\documentclass[twocolumn]{aastex631}

\usepackage{bm}

\begin{document}

\title{Excitation of post-AGB Star Binary Eccentricity by Massive Polar-Aligned Circumbinary Disks}

\author[0000-0002-7276-3694]{Shunquan Huang}
\affiliation{Department of Physics and Astronomy, University of Nevada, Las Vegas, 4505 South Maryland Parkway, Las Vegas, NV 89154, USA}
\affiliation{Nevada Center for Astrophysics, University of Nevada, Las Vegas, 4505 South Maryland Parkway, Las Vegas, NV 89154, USA}

\author[0000-0003-2401-7168]{Rebecca G. Martin}
\affiliation{Department of Physics and Astronomy, University of Nevada, Las Vegas, 4505 South Maryland Parkway, Las Vegas, NV 89154, USA}
\affiliation{Nevada Center for Astrophysics, University of Nevada, Las Vegas, 4505 South Maryland Parkway, Las Vegas, NV 89154, USA}

\author[0000-0002-4636-7348]{Stephen H. Lubow}
\affiliation{Space Telescope Science Institute, 3700 San Martin Drive, Baltimore, MD 21218, USA}



\begin{abstract}
Many post-AGB star binaries are observed to have relatively high orbital eccentricities (up to 0.6). 
Recently, AC Her was observed to have a polar-aligned circumbinary disk. 
We perform hydrodynamic simulations to explore the impact of a polar-aligned disk on the eccentricity of a binary. 
For a binary system with central masses of 0.73 $M_{\rm \odot}$ and 1.4 $M_{\rm \odot}$, we find that a disk with a total mass of 0.1 $M_{\rm \odot}$ can enhance the binary eccentricity from 0.2 to 0.7 within 5000 years, or from 0.01 to 0.65 within 15000 years.
Even if the disk mass is as low as 0.01 $M_{\rm \odot}$, the binary eccentricity grows within our simulation time while the system remains stable. 
These eccentricity variations are associated with the variations of the inclination between the disk and the binary orbit due to von Zeipel-Kozai-Lidov oscillations. 
The oscillations eventually damp and leave the binary eccentricity at a high value. 
The numerical results are in good agreement with  analytical estimates. 
In addition, we examine  the AC-Her system and  find that the disk mass should be on the order of $10^{-3}\,M_{\rm \odot}$ for the disk to remain polar. 

\end{abstract}

\keywords{Accretion -- Stellar accretion disks -- Post-asymptotic giant branch stars -- Binary stars}


\section{Introduction} \label{sec:intro}
Many post–asymptotic giant branch (AGB) stars host a stable Keplerian disk (\citealt{2006A&A...448..641D,2013A&A...557L..11B,2021A&A...648A..93G}). 
All the post-AGB stars with disks are found in binaries and the disks are all circumbinary disks (\citealt{2013A&A...557L..11B,2018Galax...6...97I}).
The orbital periods of the binaries range from tens to thousands of days (\citealt{2018A&A...620A..85O}). 
For the binaries with periods shorter than $1000\, \rm d$, it is likely that they have undergone a common envelope phase (\citealt{2012IAUS..283...95I}). 
During this phase, the binary transfers orbital energy and angular momentum to the envelope due to tidal dissipation and friction (\citealt{1993PASP..105.1373I}). 
This interaction causes the envelope to be ejected, forming a circumbinary disk (\citealt{2011MNRAS.417.1466K,2017MNRAS.468.4465C,2023MNRAS.521...35I}), while the binary orbit becomes circularized (\citealt{1977A&A....57..383Z,1981ARA&A..19..277S}).

This process is challenged by the discovery of many post-AGB binaries with high eccentricities, some as large as $0.4$ and a few even up to $0.6$ (\citealt{2018A&A...620A..85O}).
It has been suggested that such high eccentricity could be driven by gravitational torques associated with the accreting gas 
(\citealt{2007ApJ...667.1170S, 2013A&A...551A..50D, 2016ApJ...830....8R, 2019ApJ...871...84M, 2020A&A...642A.234O, 2021ApJ...914L..21D, 2021ApJ...909L..13Z}). 
However, the models in these works have their limitations, such as most of them require a massive long-lived circumbinary disk (\citealt{2013A&A...551A..50D, 2016ApJ...830....8R}). 
Moreover, all of these works assume a coplanar circumbinary disk. 
Many recent works have revealed that the circumbinary disk could be stable in a high inclination with respect to the binary if the binary is in a non-circular orbit (\citealt{Aly2015,2017ApJ...835L..28M, 2018MNRAS.479.1297M, Lubow2018,Zanazzi2018,2019MNRAS.490.1332M, 2019MNRAS.486.2919S,Cuello2019}).
In this work, we explore the impact of a polar-aligned disk on binary eccentricity, providing a new perspective on the mechanisms driving high eccentricities. 

Observations suggest that misalignments may be common between a binary orbit and its circumbinary disk (e.g., \citealt{2004ApJ...607..913C, 2004ApJ...603L..45W, 2012ApJ...757L..18C, 2012MNRAS.421.2264K, 2016ApJ...830L..16B, 2018MNRAS.480.4738A, 2018AJ....155...47A, 2019ApJ...883...22C, 2022A&A...666A..61K}). 
One interesting case is the AC-Her system, which hosts a polar-aligned circumbinary disk and potentially a polar-aligned planet (\citealt{2015A&A...578A..40H, 2021A&A...648A..93G, 2023ApJ...950..149A, 2023ApJ...957L..28M}). 
While the binary orbit is well determined, the total disk mass and size remain debated. 
These features are crucial since a polar-aligned circumbinary disk is stable only when the disk mass, and thus the angular momentum ratio between the disk and binary, is low (\citealt{2017ApJ...835L..28M, 2019MNRAS.490.1332M,2019MNRAS.490.5634C}).
If the disk is significantly radially extended, its outer regions provide substantial angular momentum. This can trigger von Zeipel-Kozai-Lidov (ZKL; \citealt{1910AN....183..345V,1962AJ.....67..591K,1962P&SS....9..719L}) oscillations of the binary in which the binary eccentricity and inclination oscillate \citep[e.g.][]{Terquem2010,Lepp2023,Martin2024}.
In this paper, we discuss how the mass and size of a polar-aligned disk affect its stability and aim to establish a mass limit for the AC-Her system. Our model explores the role of the circumbinary disk on the evolution of the non-coplanar binary. We do not include effects that could arise in the gap region or in the circumstellar disks \citep[see e.g.][]{2019ApJ...871...84M,2021ApJ...909L..13Z}. 

This paper is structured as follows: we describe the simulation setup in Section~\ref{sec:setup}. 
In Section~\ref{sec:ecc_excitation}, we explore the evolution of a polar-aligned circumbinary disk around an eccentric binary using hydrodynamic simulations.
In Section~\ref{sec:ACHer}, we examine the disk features of the AC-Her system. 
Finally, we draw our conclusions in Section~\ref{sec:conclusion}.

\section{Problem Setup}\label{sec:setup}
To explore the effects of polar-aligned circumbinary disks on the binary, we run hydrodynamic simulations using the smoothed particle hydrodynamics (SPH; e.g., \citealt{2012JCoPh.231..759P}) code {\sc phantom} (\citealt{2010MNRAS.405.1212L,2010MNRAS.406.1659P,2018PASA...35...31P}). 
{\sc phantom} has been widely used for simulations of misaligned accretion disks in binary systems (e.g., \citealt{2012MNRAS.423.2597N, 2013MNRAS.434.1946N, Facchini2013,2014ApJ...792L..33M, 2015ApJ...807...75F,Franchini2019,Nealon2020}). 
We set up the simulation based on the observational results of the AC Her system (\citealt{2015A&A...578A..40H, 2023ApJ...950..149A, 2023ApJ...957L..28M}). The units we use for mass, time, and length are $M_\odot$, years, and $\rm AU$, respectively.
The masses of the post-AGB star and the companion star in the AC Her system are $M_1 = 0.73\,{\rm M_\odot}$ and $M_2 = 1.4\,{\rm M_\odot}$, respectively. 
The binary semi-major axis is $a_{\rm b} = 2.83\,{\rm AU}$, and the eccentricity is $e=0.206$. 
The inclination of the circumbinary disk with respect to the binary is observed to be $i_{\rm bd} = 96.^{\circ}5$ with error bars that allow for a polar-aligned disk ($i_{\rm bd}=90^\circ$). 
In our simulations, the binary stars are treated as two sink particles each with a sink radius of $0.5\,\text{AU}$.  Any gas particle within $0.5\,\text{AU}$ of either star is directly accreted by the star. The mass and angular momentum of an accreted particle are added to the sink particle \citep{Bate1995}.
This is a relatively large sink radius since we focus on the circumbinary disk and are not simulating the small disks around each star (see \citealt{1996ApJ...467L..77A, 2019ApJ...871...84M, 2020ApJ...889..114M} for more details).

We assume that the disk is initially polar-aligned to the binary, i.e., $i_{\rm bd} = 90^{\circ}$ with the disk angular momentum vector aligned to the binary eccentricity vector.
The disk initially extends from $r_{\rm in} = 9\,{\rm AU}$ to $r_{\rm out} = 50\,{\rm AU}$ with a radial density profile following a power law $\Sigma = \Sigma_0r^{-p}$ where the index $p=1.5$ and $\Sigma_0$ is defined by the total disk mass. 
The disk expands inwards and outwards over time. The inner parts of the disc reach a quasi-equilibrium state within a few binary orbits, while our simulations run for $6000$ binary orbits.

Other parameters and resolutions for our simulations are listed in Table~\ref{tab:simpara}. 
The initial number of particles $n_p$ represents the resolution, and $H/r$ is the scale height of the disk as a ratio of the radius following $H/r = c_s/v_\phi$ where $c_s$ is the sound speed following $c_s\propto r^{-q}$ with $q=0.75$, and $v_\phi$ is the Keplerian velocity following $v_\phi\propto r^{-\frac12}$. 
Thus, $H/r$ decreases weakly  with radius following $H/r\propto r^{-0.25}$. The values of $H/r$ in Table~\ref{tab:simpara} are at  the initial inner radius of $r_{\rm in} = 9\,{\rm AU}$.
We adopt an isothermal equation of state.  The disk is initially set with a slightly sub-Keplerian velocity that includes a pressure correction.
The viscosity is prescribed using the equivalent \cite{1973A&A....24..337S} $\alpha$  disc viscosity based on Equation~38 in \cite{2010MNRAS.405.1212L}, 
\begin{equation}
    \alpha \approx \frac{1}{10}\alpha^{\rm AV} \frac{\left<h\right>}{H}, 
\end{equation}
where $\alpha^{\rm AV}$ is the artificial viscosity in {\sc phantom}, and $\left<h\right>$ is the azimuthally averaged smoothing length. 
When the viscosity $\alpha$ is initially specified and the mean smoothing length $\left<h\right>$ is determined by the number of particles, the artificial viscosity $\alpha^{\rm AV}$ is calculated automatically in {\sc phantom} to give the desired viscosity.

\begin{table*}[t]
        \begin{center}
	\caption{Parameters for simulations in this work. From left to right, the quantities are as follows: simulation name, number of particles $n_p$, disk mass $M_{\rm \rm disk}$, binary eccentricity $e_{\rm 0}$, \cite{1973A&A....24..337S} disk viscosity $\alpha$, disk aspect ratio $H/r$ at $r=9\,\rm AU$, artificial viscosity $\alpha^{\rm AV}$, and the ratio between the azimuthally averaged smoothing length and the disk scale-height $\left<h\right>/H$.}  
	\label{tab:simpara} 
	\begin{tabular}{cccccccc}
        \toprule
        Name   &  $(n_p)$ &  $M_{\rm \rm disk}(M_{\rm \odot})$ & $e_{\rm 0}$ & $\alpha$ &  H/r &  $\alpha^{\rm AV}$ &  $\left<h\right>/H$ \\
        \hline
	run 1   &   $1.0\times10^6$  &    0.01  &  0.206  &   0.01 &  0.1 & 0.446 & 0.224 \\
	run 2   &   $1.0\times10^6$  &    0.05  &  0.206  &   0.01 &  0.1 & 0.446 & 0.224 \\
	run 3   &   $1.0\times10^6$  &    0.10  &  0.206  &   0.01 &  0.1 & 0.446 & 0.224 \\
        run 4   &   $3.0\times10^5$  &    0.01  &  0.206  &   0.01 &  0.1 & 0.299 & 0.335 \\
        run 5   &   $3.0\times10^5$  &    0.05  &  0.206  &   0.01 &  0.1 & 0.299 & 0.335 \\
        run 6   &   $3.0\times10^5$  &    0.10  &  0.206  &   0.01 &  0.1 & 0.299 & 0.335 \\
        run 7   &   $1.0\times10^5$  &    0.01  &  0.206  &   0.01 &  0.1 & 0.207 & 0.483 \\
        run 8   &   $1.0\times10^5$  &    0.05  &  0.206  &   0.01 &  0.1 & 0.207 & 0.483 \\
        run 9   &   $1.0\times10^5$  &    0.10  &  0.206  &   0.01 &  0.1 & 0.207 & 0.483 \\
        run 10  &   $3.0\times10^5$  &    0.10  &  0.206  &   0.01 &  0.05 & 0.188 & 0.532 \\ 
        run 11  &   $3.0\times10^5$  &    0.10  &  0.206  &   0.10 &  0.1 & 2.987  & 0.335 \\ 
        run 12  &   $3.0\times10^5$  &    $8.1\times10^{-4}$  &  0.206  &   0.01 &  0.1 & 0.299 & 0.335 \\
        run 13  &   $3.0\times10^5$  &    0.01  &  0.010  &   0.01 &  0.1 & 0.299  & 0.335 \\
        run 14  &   $3.0\times10^5$  &    0.05  &  0.010  &   0.01 &  0.1 & 0.299  & 0.335 \\
        run 15  &   $3.0\times10^5$  &    0.10  &  0.010  &   0.01 &  0.1 & 0.299  & 0.335 \\
        run 16  &   $3.0\times10^5$  &    0.01  &  0.050  &   0.01 &  0.1 & 0.299  & 0.335 \\
        run 17  &   $3.0\times10^5$  &    0.05  &  0.050  &   0.01 &  0.1 & 0.299  & 0.335 \\
        run 18  &   $3.0\times10^5$  &    0.10  &  0.050  &   0.01 &  0.1 & 0.299  & 0.335 \\
        
        \hline
	\end{tabular}
    \end{center}
\end{table*}

\section{Interaction between binary and polar-aligned disk}\label{sec:ecc_excitation}

In this section we describe the results of the simulations runs 1-9 that have various disk masses and resolutions as shown in Table~\ref{tab:simpara}.
The results from these simulations reveal the evolution of the binary interacting with a polar-aligned disk. 

\subsection{Evolution of the binary eccentricity}
\begin{figure}
    \includegraphics[width=\columnwidth]{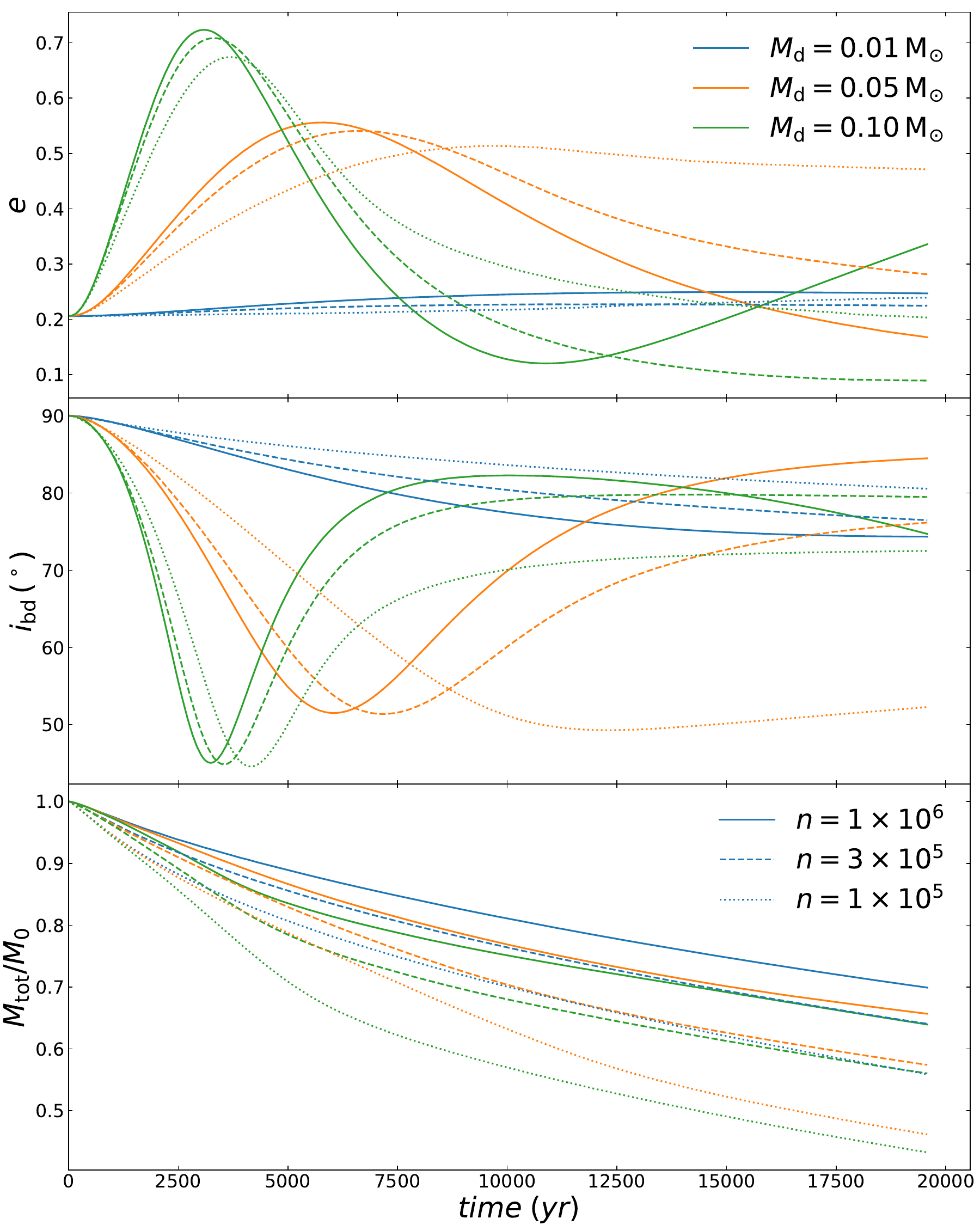}
    \caption{Features of the binary and the disk as functions of time for simulations run 1-9. 
    The upper, middle, and bottom panels show the binary eccentricity, disk inclination with respect to the binary, and the normalized disk mass, respectively. 
    The blue, orange, and green lines represent the initial total disk mass $M_{\rm d}$ of $0.01$, $0.05$, and $0.1\,{\rm M_\odot}$, respectively.  
    The solid, dashed, and dot lines represent the resolution $n_p = 1\times10^6$, $3\times10^5$, and $1\times10^5$, respectively. 
    All simulations shown here use a viscosity $\alpha = 0.01$. 
    }
    \label{fig:ecc_inc_mass}
\end{figure}

The first crucial question to answer is how a polar-aligned disk affects the binary's eccentricity.  
The upper panel of Figure~\ref{fig:ecc_inc_mass} shows the binary eccentricity over time for various disk masses and resolutions.

We first examine the highest resolution cases (i.e., $n_p=1.0\times 10^6$, run~1-3). 
If the disk has a mass of $0.1\,{\rm M_\odot}$ (run~3), the eccentricity grows above 0.7 in $3\times 10^4$ years and then undergoes damped oscillations. 
These eccentricity oscillations are correlated with disk inclination oscillations with respect to the binary (see Section~\ref{sec:disk_inc}), due to the ZKL oscillations.
For the case with an initial mass of $0.05\,{\rm M_\odot}$ (run~2), the eccentricity also rises to about $0.55$ but over a longer timescale.
The ZKL oscillations depend on the disk's mass, or more specifically, its angular momentum. 
As the disk loses mass, the oscillations diminish in both amplitude and period.

For the lower resolution cases (run~4-9), the oscillations damp more quickly due to larger numerical viscosity and faster disk mass loss (see Section~\ref{sec:diskmass}.) 
This damping effect is also observed in \cite{2017ApJ...835L..28M}. 
For instance, when $M_{\rm d}=0.05\,{\rm M_\odot}$, the low-resolution case ($n=1\times10^5$, dotted line) shows oscillations quenching at about 10000 years, with eccentricity remaining at $0.5$. 
In the medium-resolution case ($n=3\times10^5$, dashed line), the oscillations stop close to the end of the simulation, with a final eccentricity of about $0.3$. 
For the high-resolution case, the evolution continues beyond our simulation timeframe. However, we can presume that the eccentricity will eventually stabilize at a potentially high value once the disk loses sufficient mass.

\subsection{Evolution of the Circumbinary Disk's Inclination} \label{sec:disk_inc}
The inclinations of the disk with respect to the binary over time are shown in the middle panel of Figure~\ref{fig:ecc_inc_mass}. 
The disk inclination with respect to the binary $i_{\rm bd}$ is calculated with
\begin{equation}
    \cos i_{\rm bd} = \frac{\bm{L}_{\rm disk} \cdot \bm{L}_{\rm bin}}
                           {|\bm{L}_{\rm disk}||\bm{L}_{\rm bin}|}  \,,
    \label{eq:i_bd}
\end{equation}
where $\bm{L}_{\rm disk}$ and $\bm{L}_{\rm bin}$ are the total angular momentum of the disk and binary, respectively. 
Generally, the inclinations drop dramatically at the beginning and then transition into oscillations over a longer timescale. 
The inclination oscillations with respect to the binary are highly correlated with the binary eccentricity. 
Specifically, a minimum in the inclination corresponds to a peak in eccentricity and vice versa.
Lower disk masses result in extended oscillation timescales, while lower resolutions cause the evolution to stop earlier.

\begin{figure}
    \includegraphics[width=\columnwidth]{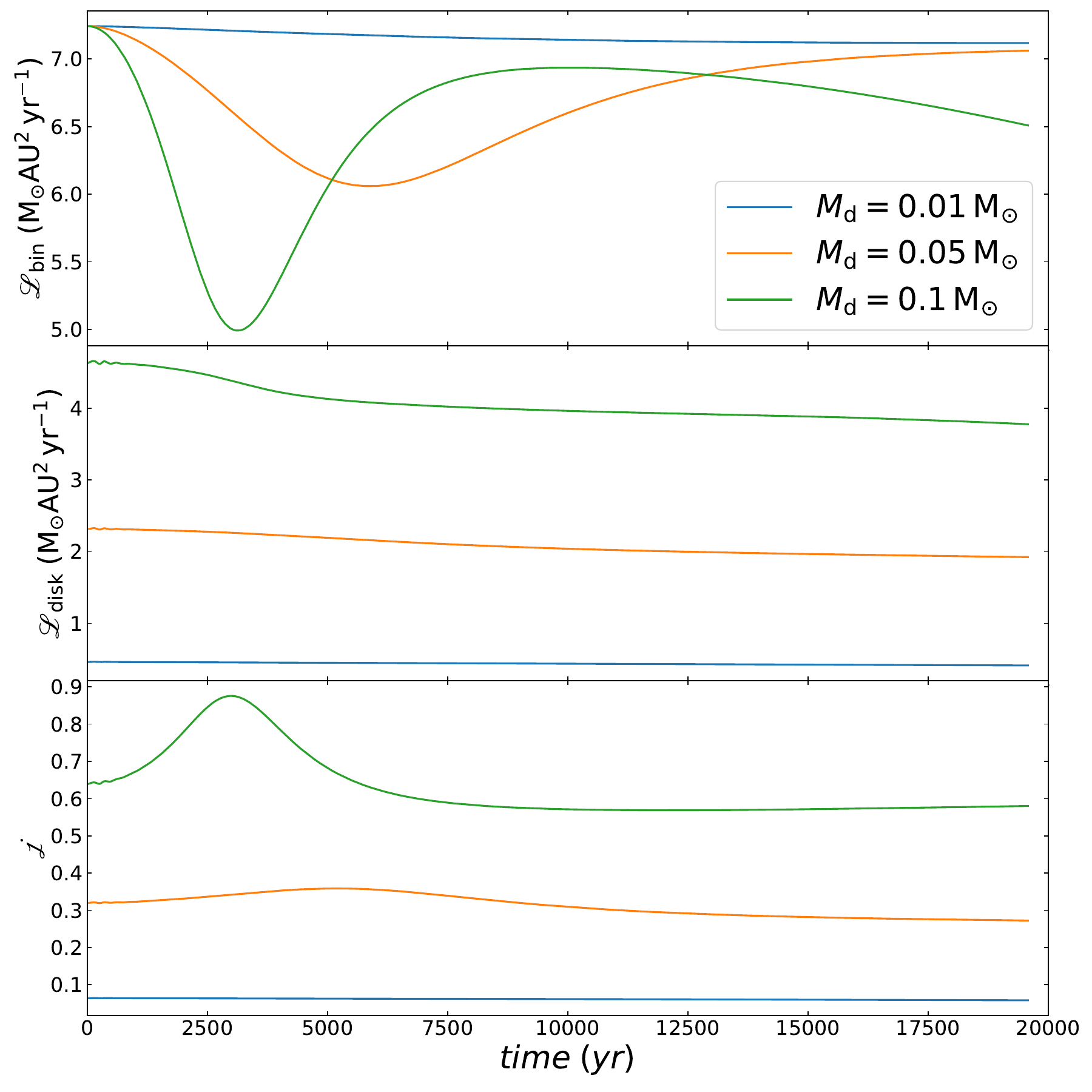}
    \caption{The angular momentum of the binary, $L_{\rm bin}$ (upper panel), the disk, $L_{\rm disk}$ (middle panel), and the ratio $j$ (bottom panel) between the disk and the binary as functions of time for simulation run 1-3.
    The blue, orange, and green lines represent the initial total disk mass $M_{\rm d}$ of $0.01$, $0.05$, and $0.1\,{\rm M_\odot}$, respectively. }
    \label{fig:jratio}
\end{figure}

The ZKL oscillations of the disk's inclination with respect to the binary and eccentricity of the binary can be understood via the disk's stationary inclination. The stationary inclination of a ring is given by
\begin{equation}
    \cos i_{\rm s} = \frac{ -(1+4e^2) + \sqrt{(1+4e^2)^2 + 60(1-e^2)j^2}}{10j}\,,
    \label{eq:StableInc}
\end{equation}
\citep{2019MNRAS.490.1332M,2019MNRAS.490.5634C}, 
where $e$ is the binary's eccentricity, and $j=L_{\rm disk}/L_{\rm bin}$ is the angular momentum ratio between the disk and the binary. 
This concept is essentially equivalent to similar derivations by \cite{2010MNRAS.401.1189F}, as well as those presented in \cite{2013MNRAS.431.2155N} and \cite{2017MNRAS.467.3066A}. 
Note that if the disk is massless, the angular momentum ratio $j$ approaches $0$, then equation~(\ref{eq:StableInc}) yields a stationary inclination $i_{\rm s}\approx 90^\circ$ with respect to the binary. 
The angular momentum of the binary $L_{\rm bin}$, the disk $L_{\rm disk}$, and the ratio $j$ from our simulation run 1-3 are shown in Figure~\ref{fig:jratio}. 
The angular momentum of the binary is $L_{\rm bin} = \mu\sqrt{GMa(1-e^2)}$, where $\mu = \frac{M_1M_2}{M_1+M_2}$ is the reduced mass of the binary and $M$ is the total mass. 
Since variations in the binary's mass and semi-major axis are negligible, changes in $L_{\rm bin}$ primarily depend on variations in the binary's eccentricity as shown in the first panel of Figure~\ref{fig:jratio}. 
The disk angular momentum, $L_{\rm disk}$, is integrated over the disk from $4.5\,{\rm AU}$ to $100\,{\rm AU}$.
In the second panel, it's important to note that mass accretion results in angular momentum loss. 
But the mass loss involves relatively low angular momentum material.
Consequently,  there is a fairly flat evolution of $L_{\rm disk}$ (see detailed discussion in Section~\ref{sec:diskmass}).
$L_{\rm disk}$ continues to decrease because we exclude a small amount of angular momentum carried by the circumstellar disks within the cavity and a few outflowing particles.
As a result, $j$ is relatively smooth overall, with a few fluctuations induced by the eccentricity of the binary. 

\begin{figure}
    \includegraphics[width=\columnwidth]{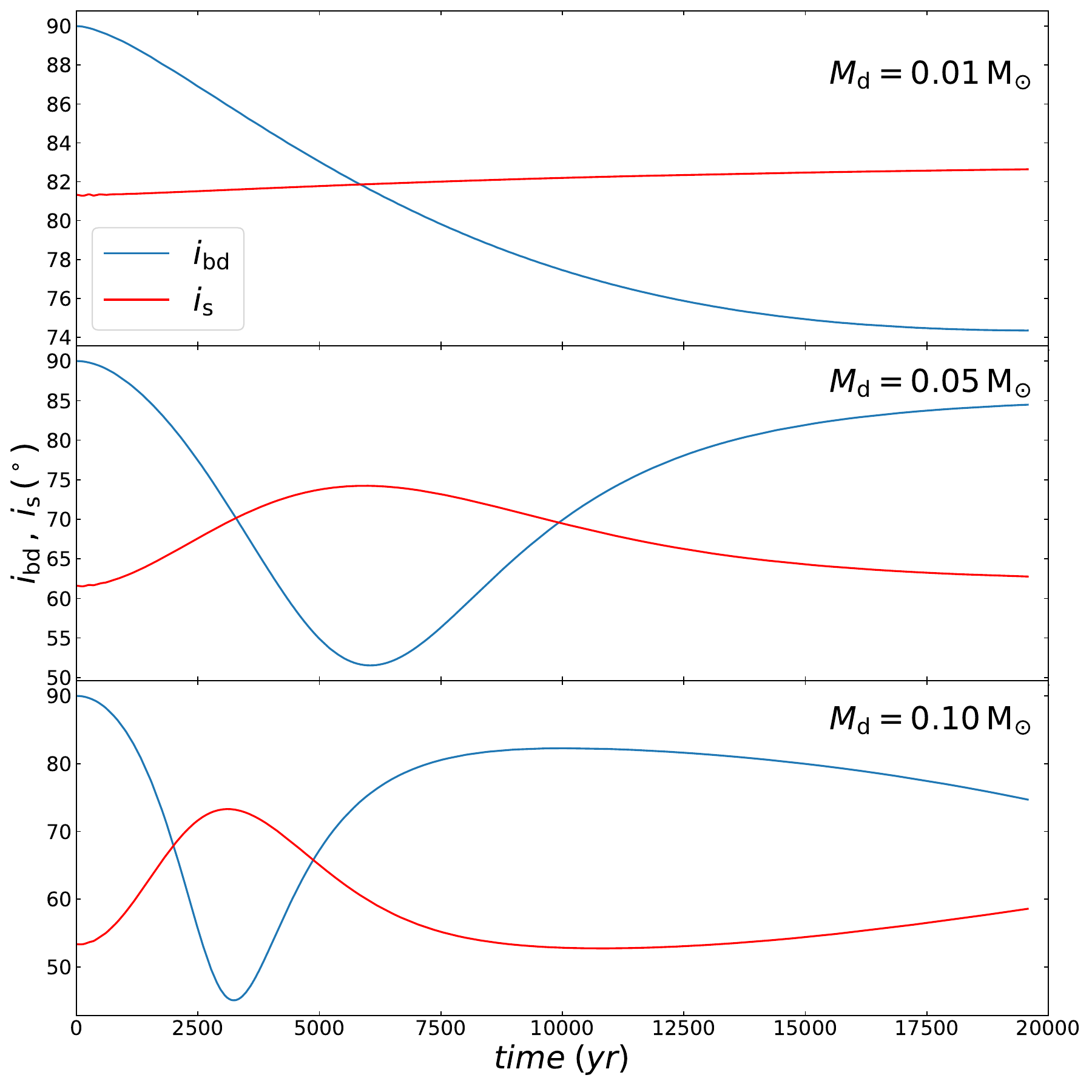}
    \caption{Disk inclination with respect to the binary ($i_{\rm bd}$) compared to the analytical stationary inclination ($i_{\rm s}$) with respect to the binary for simulation run 1-3. 
            The blue lines represent the disk inclination with respect to the binary from the simulation, and the red lines represent the stationary solutions (equation~\ref{eq:StableInc}).
            The upper, middle, and bottom panels show the initial total disk mass $M_{\rm d}$ of $0.01$, $0.05$, and $0.1\,{\rm M_\odot}$, respectively.}
    \label{fig:inc_ana}
\end{figure}

Figure~\ref{fig:inc_ana}  shows the analytical stationary inclination (equation~(\ref{eq:StableInc})) in the red lines and the highest resolution simulation results from runs 1-3 in the blue lines.
A larger disk mass results in a higher angular momentum ratio $j$, which corresponds to a lower stationary inclination of the disk relative to the binary initially. 
According to equation~(\ref{eq:StableInc}), $i_{\rm s}$ positively correlates with $e$. 
Hence, when the eccentricity $e$ of the binary peaks, so does $i_{\rm s}$.
This relationship is evident in the major features of $i_{\rm s}$ over time, as illustrated in Figure~\ref{fig:inc_ana}. 
Initially the disk's inclination with respect to the binary is $i_{\rm bd}=90^\circ$ in all simulations. 
This causes a peak difference between $i_{\rm bd}$ and $i_{\rm s}$ at the beginning, leading to a dramatic decrease in $i_{\rm bd}$.
Subsequently, $i_{\rm bd}$ oscillates about the stationary inclination $i_{\rm s}$ over time.
Therefore the oscillations have a larger amplitude for higher disk mass.
As the simulations progress, the disk's inclination with respect to the binary gradually damps due to viscous dissipation.  
Accordingly, the eccentricity of the binary tends towards a stable, relatively high value.

\subsection{Disk Mass} \label{sec:diskmass}
Most of the lost disk mass is accreted onto the central objects. 
Therefore, the disk's mass loss rate also correlates with the accretion rate onto the binary. 
In the bottom of Figure~\ref{fig:ecc_inc_mass}, we show the disks' mass normalized by their initial mass as functions of time. 
The total disk mass is calculated by the mass between $4.5\,{\rm AU}$ and $100\,{\rm AU}$. 
From Figure~\ref{fig:ecc_inc_mass}, we see that the low-resolution cases lose mass much faster than the high-resolution cases. 
For instance, the $M_{\rm d}=0.1\,{\rm M_\odot}$ case loses more than half of the disk mass by the end of the simulation for low-resolution (green dot line), but only $35\%$ for high-resolution (green solid line). 

\begin{figure}
    \includegraphics[width=\columnwidth]{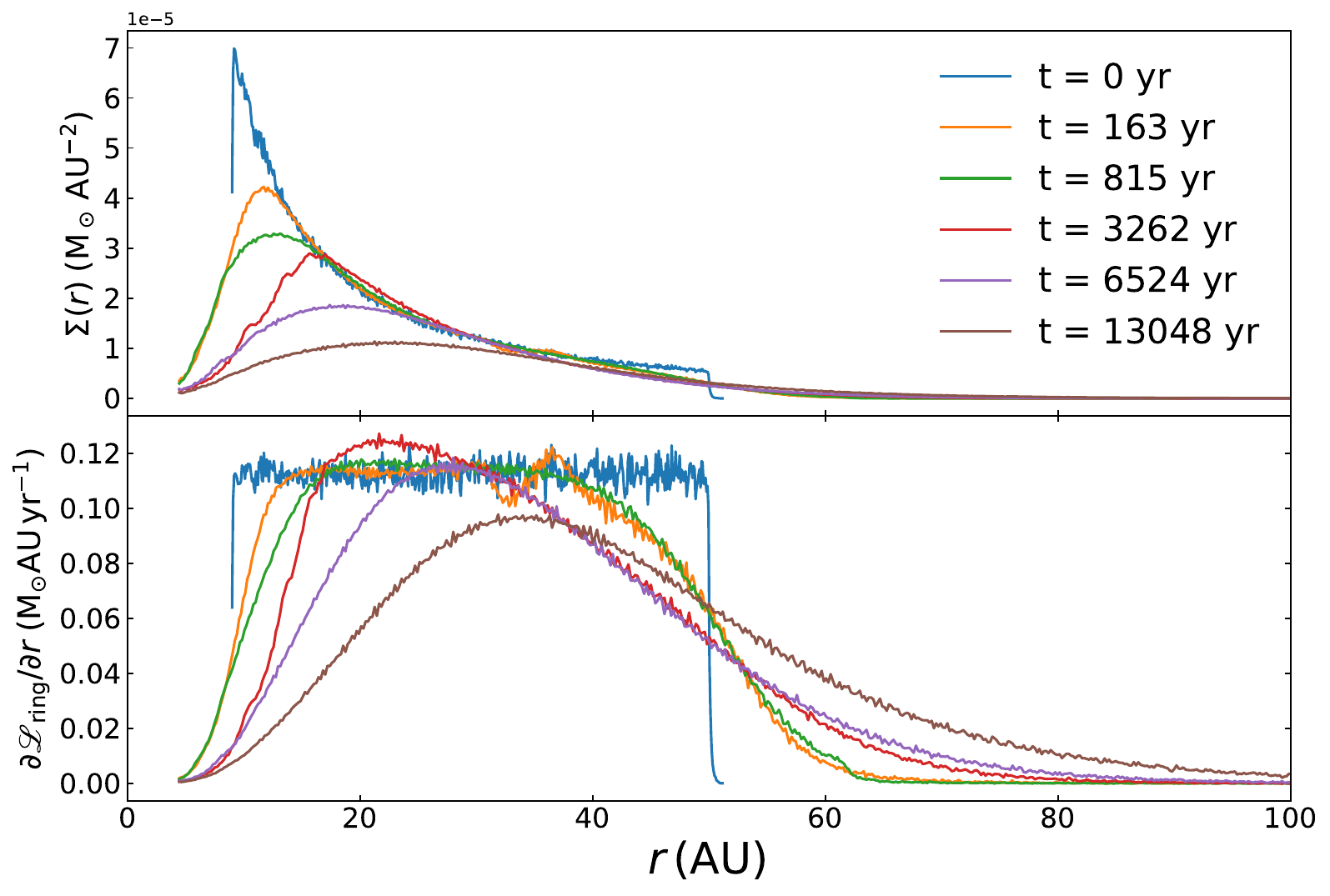}
    \caption{
             Disk mass and angular momentum distribution at different times for simulation run~3. 
             The  surface density $\Sigma(r)$ is shown in the upper panel, while the azimuthal angular momentum as a function of $r$ is presented in the lower panel. 
             Different colors correspond to different times. 
             }
    \label{fig:sigma_profile}
\end{figure}
The disk mass and angular momentum distributions at different times for $M_{\rm d}=0.1\,{\rm M_\odot}$ (high resolution, run~$3$) are shown in Figure~\ref{fig:sigma_profile}. 
From the surface density $\Sigma$ in the upper panel, we can see that the initially sharp truncation of the disk smooths out over a relatively short timescale, forming a relaxed profile with a high peak density well before the onset of ZKL oscillations.
The material in the inner disk is accreted, causing the surface density in the inner disk to decrease over the long-term evolution.
Since the total disk angular momentum is approximately conserved, angular momentum from the inner disk is transported outwards, as shown in the lower panel of Figure~\ref{fig:sigma_profile}. 
The angular momentum distribution is given by $\partial \mathcal{L}_{\rm ring}/\partial r = v_\phi r dm/dr$ where $v_\phi$ is the azimuthal velocity and mass $dm=\Sigma(r)\cdot 2\pi r dr$.
This outward transport of angular momentum increases the interaction timescale between the disk and the binary. 
As a result, the ZKL oscillation timescale also becomes longer, as illustrated in Figure~\ref{fig:ecc_inc_mass} above. 

\begin{figure}[htb!]
    \includegraphics[width=\columnwidth]{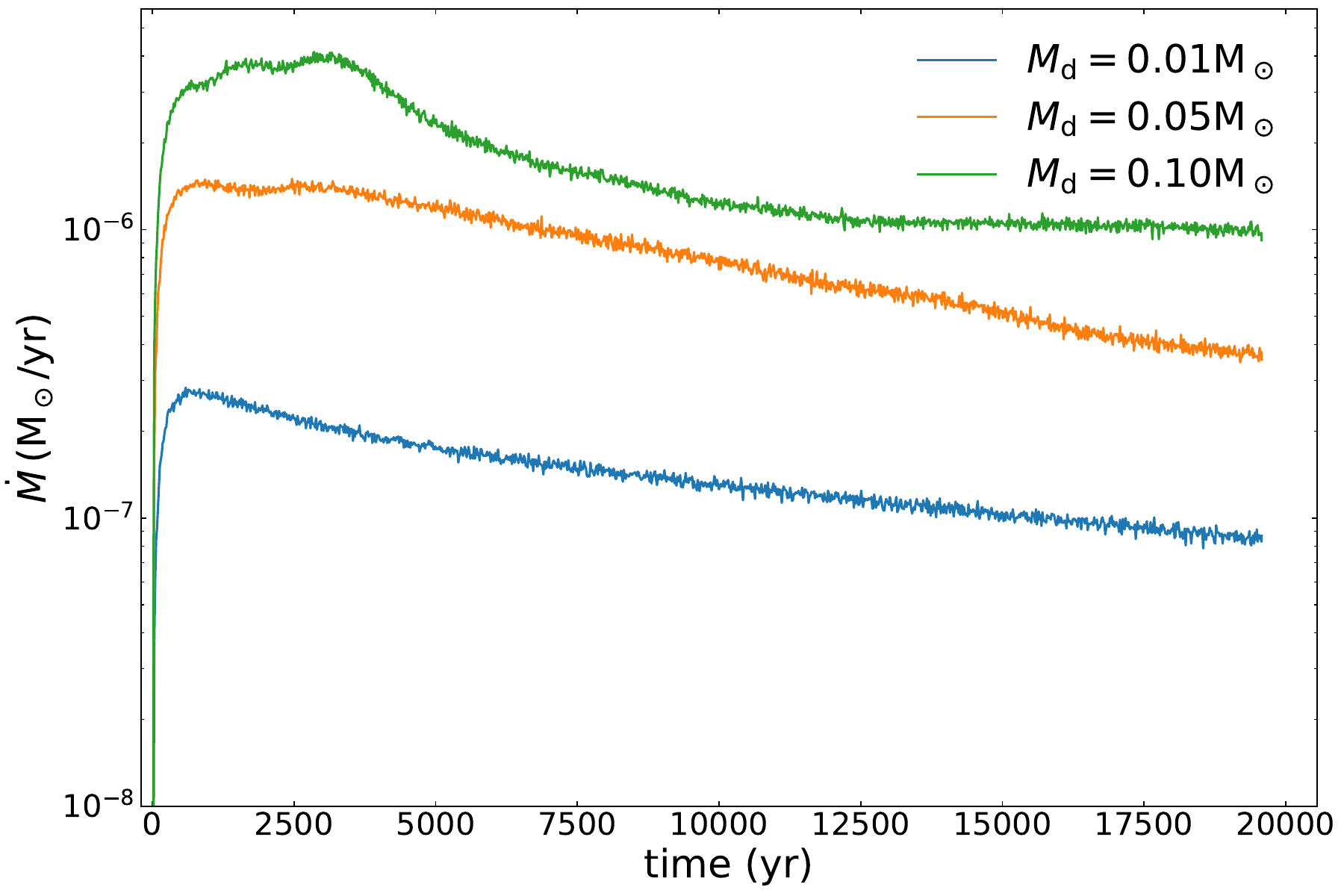}
    \caption{
             Total mass accretion rate on both stars as a function of time for simulation run~1-3.
             The accretion rate is measured directly from the mass increases of the stars.
             The blue, orange, and green lines represent the initial total disk mass $M_{\rm d}$ of $0.01$, $0.05$, and $0.1\,{\rm M_\odot}$, respectively.
             }
    \label{fig:mdot}
\end{figure}
When considering the same resolution, the disk's mass evolution scaled to the initial disk mass is almost identical between different initial masses, e.g., the high-resolution cases (run~1-3) all have a mean normalized accretion rate $\dot{M}/M_{\rm d0} \sim 10^{-5} \,\rm yr^{-1}$. 
However, since the normalized factors differ for these cases, they have different accretion rates.
As we can see from Figure~\ref{fig:mdot}, the accretion rate is about $\dot{M}\sim 10^{-7} \rm M_\odot\,yr^{-1}$ for $M_{\rm d}=0.01\,{\rm M_\odot}$ (run~1), and $\dot{M}\sim 10^{-6} \rm M_\odot\,yr^{-1}$ for $M_{\rm d}=0.1\,{\rm M_\odot}$ (run~3). 
Nevertheless, it is still challenging to constrain the accretion rate based on the simulations because it depends on multiple factors, such as the viscosity (see \ref{sec:vis}), and the sink radius of the binary in SPH simulation, etc.

\subsection{Disk Viscosity}\label{sec:vis}
\begin{figure}[htb!]
    \includegraphics[width=\columnwidth]{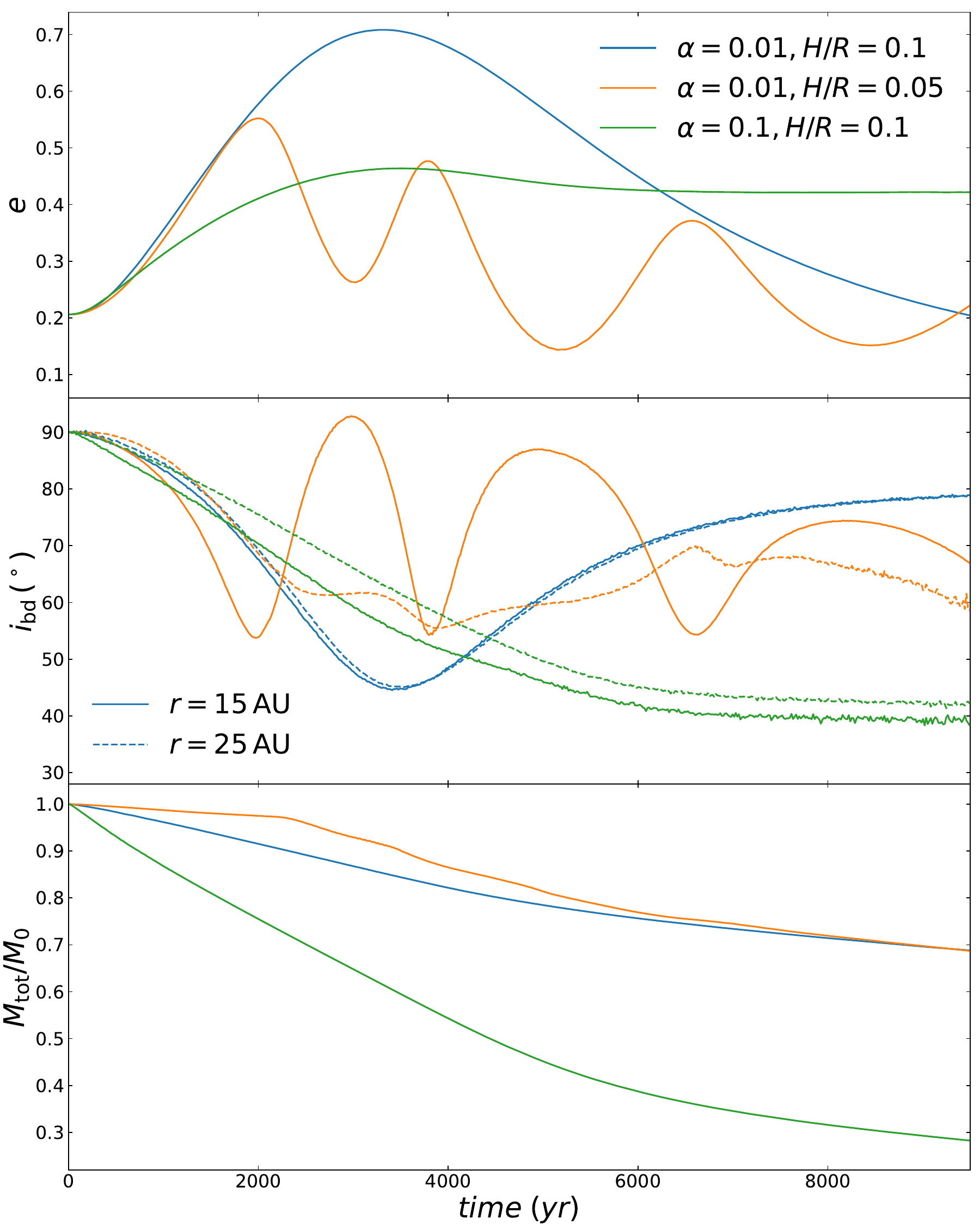}
    \caption{Binary eccentricity (upper panel), disk's inclinations with respect to the binary (middle panel), and disk mass (bottom panel) as functions of time for simulation run~6, 10, and 11.    
    The blue lines represent the case for $\alpha = 0.01$ and $H/r=0.1$, orange lines for $\alpha = 0.01$ and $H/r=0.05$, and green lines for $\alpha = 0.1$ and $H/r=0.1$. 
    In the middle panel, the solid lines and dashed lines represent the disk's inclinations relative to the binary at $r=15\,{\rm AU}$ and $r=25\,{\rm AU}$.
    All simulations presented here were carried out with a resolution of $n=3\times 10^5.$
             }
    \label{fig:diffv}
\end{figure}
The disk dissipation depends essentially on the viscosity. 
Given the close connection between the final state of the binary eccentricity and the disk, it is of significant interest to explore how the disk behaves under different viscosity conditions. 
To explore the effects of different viscosity levels, we use simulation run~6 in Table~\ref{tab:simpara} as the baseline and modify it to setup run~10 and 11 for lower and higher viscosity, respectively.
For the higher viscosity case (run~11), we use the same setup as run~6 but increase the disk viscosity parameter to $\alpha=0.10$. 
For the lower viscosity case (run~10), directly reducing $\alpha$ in SPH simulations can introduce numerical challenges.
Instead, we reduce the disk scale height $H$ to achieve a lower effective viscosity in the simulation, as the kinematic viscosity is given by $\nu = \alpha c_s H$. 
The binary eccentricity, disk inclination relative to the binary, and the disk mass evolutions are shown in Figure~\ref{fig:diffv}. 

In Figure~\ref{fig:diffv}, for the high-viscosity case (green line, run~11), the rapid dissipation of the disk prevents the binary eccentricity from reaching the same maximum value observed in the standard case (blue line, run~6).
Even so, we might gain insight into how the ZKL oscillation ends.
As expected, the binary's eccentricity remains essentially constant after the disk loses most of its mass. 

\begin{figure}
    \includegraphics[width=\columnwidth]{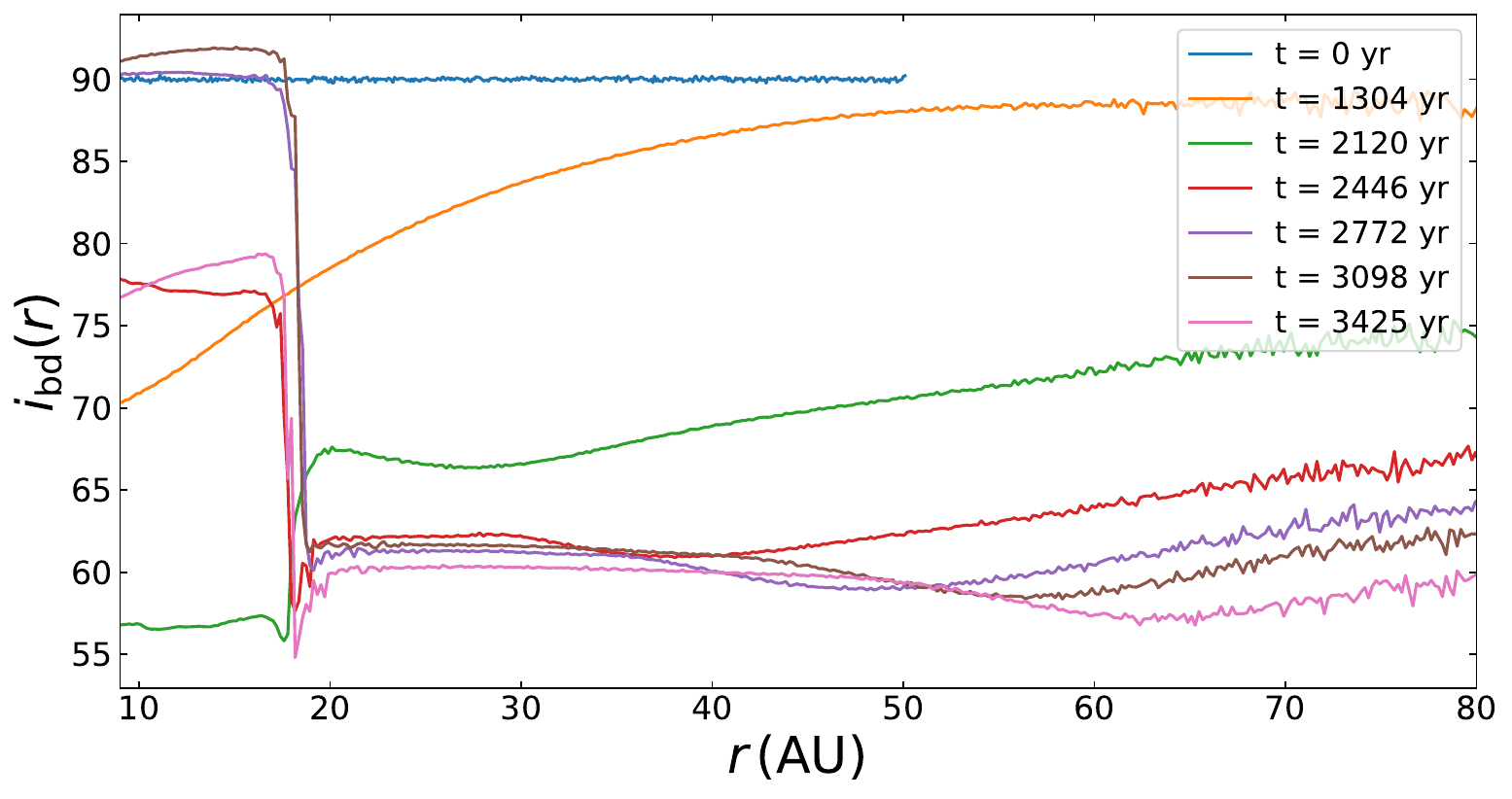}
    \caption{
             The disk inclinations with respect to the binary as functions of the radius at different times for simulation run~10. 
             Different colors represent different time, respectively. 
             }
    \label{fig:i_bd_disk_break}
\end{figure}

For the low-viscosity case (run~10), the initial mass loss is much lower. 
However, the accretion rate undergoes a transition at a time of about $t=2000$ years, which is due to the disk breaking.   
When the viscosity is low, the circumbinary disk inclination varies with radius due to the strong gravitational torques from the binary. 
The communication timescale through the disc is longer than the precession timescale and the disk breaks into two disjoint sections with different inclinations. 
As shown in Figure~\ref{fig:i_bd_disk_break}, the disk breaks at a radius of $r\approx 20\,{\rm AU}$ (see e.g. \citealt{1983MNRAS.202.1181P,1995ApJ...438..841P,1999MNRAS.304..557O,2010MNRAS.405.1212L,Nixon2012,Nixon2016,Nealon2016,Young2023,2024MNRAS.tmp.1767R} for discussions on disk warping and breaking).
Following the break, the outer disk remains almost fixed, while the inner disk undergoes faster precession, as depicted in the middle panel of Figure~\ref{fig:diffv} (orange lines). 
Since the inner disk is closer to the binary, its effect dominates the binary evolution. 
Consequently, the binary's eccentricity oscillates on a shorter timescale.

\subsection{Initially low eccentricity}
We have so far demonstrated that the binary's eccentricity can significantly grow from an initial value of $0.2$. 
However, \cite{2018A&A...620A..85O} indicates that many systems stabilize at an eccentricity of around $0.2$. 
To investigate whether the binary's eccentricity can grow from lower initial values, we conduct simulations (run $13-18$) with initial binary eccentricities of $e_{\rm 0}=0.01$ and $0.05$. 

\begin{figure}[htb!]
    \includegraphics[width=\columnwidth]{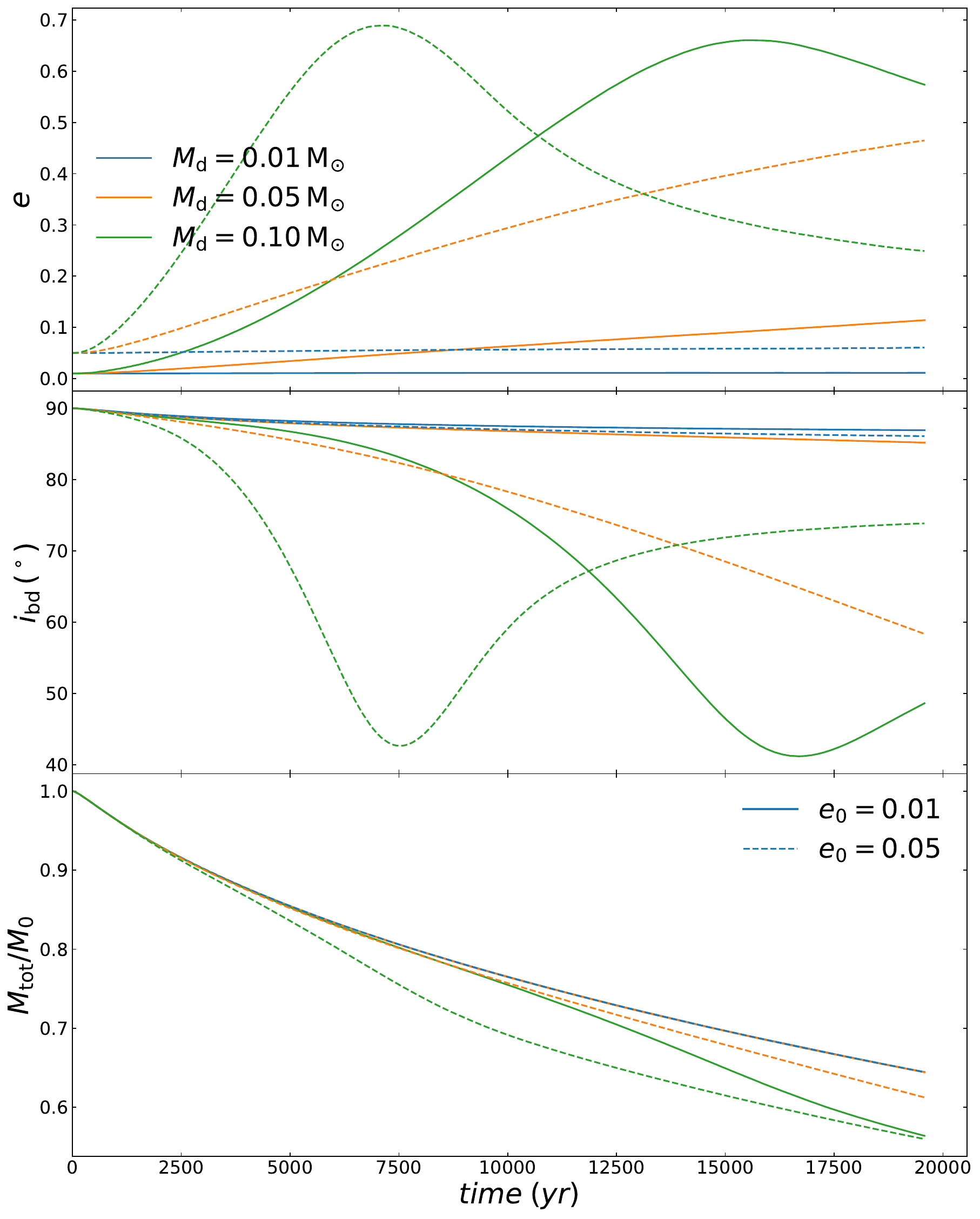}
    \caption{Features of the binary and the disk as functions of time for simulations run 13-18. 
    The upper, middle, and bottom panels show the binary eccentricity, disk inclination with respect to the binary, and the normalized disk mass, respectively. 
    The blue, orange, and green lines represent the initial total disk mass $M_{\rm d}=0.01$, $0.05$, and $0.1\,{\rm M_\odot}$, respectively.  
    The solid and dashed lines represent the initial binary's eccentricity $e_{\rm 0}=0.01$ and $0.05$, respectively. 
    All simulations presented here were carried out with a resolution of $n=3\times 10^5.$
    }
    \label{fig:lowe}
\end{figure}

As shown in Figure~\ref{fig:lowe}, the results indicate that the binary's eccentricity can still grow significantly when the disk mass is large, although the evolution timescale becomes longer for lower initial eccentricities. 
Notably, when $M_{\rm d}=0.05 \,{\rm M_{\odot}}$ and $e_0=0.01$ (run~14), the binary's eccentricity grows to $0.1$ in $20000$ years, while the disk's inclination relative to the binary changes by only about $5^\circ$ from polar alignment. 
This result may provide an explanation for binary systems with eccentricities in the range of $0.1\sim0.2$, as discussed in \cite{2018A&A...620A..85O}. 

\begin{figure}[htb!]
    \includegraphics[width=\columnwidth]{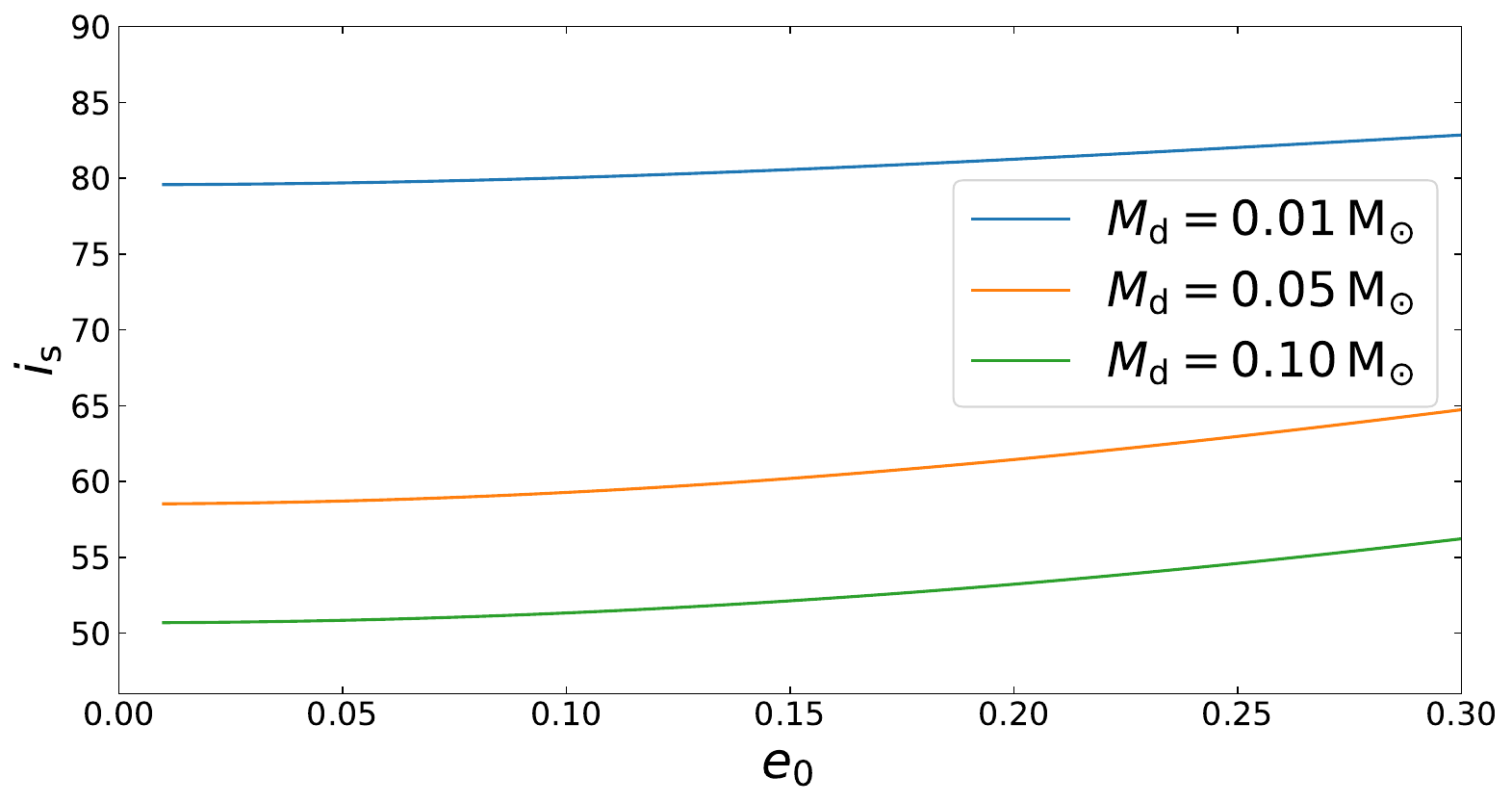}
    \caption{Stationary inclination between the disk and the binary as a function of eccentricity in the initial state, following Equation~\ref{eq:StableInc}, where $j$ is the angular momentum ratio between the initial disk and the binary. 
    The blue, orange, and green lines represent the initial total disk mass $M_{\rm d}=0.01$, $0.05$, and $0.1\,{\rm M_\odot}$, respectively.  
    }
    \label{fig:i_ana_e}
\end{figure}

The binary eccentricity growth from an initially low value can be understood by examining the initial disk's stationary inclination $i_{\rm s}$, which depends on the initial eccentricity $e_0$, as shown in Figure~\ref{fig:i_ana_e}.
When the disk's mass is large, $i_{\rm s}$ deviates from $90^\circ$ even for low binary eccentricities. 
Consequently, if the disk is initially polar-aligned with respect to the binary, the system becomes dynamically unstable, triggering ZKL oscillations. 
These oscillations cause the disk's inclination relative to the binary to decrease, while the binary's eccentricity increases.

\section{Disk features around AC Her}\label{sec:ACHer}
Previous works have provided substantial constraints on the binary and the disk in the AC Her system, except the disk's mass and radial extent (\citealt{2015A&A...578A..40H, 2021A&A...648A..93G, 2023ApJ...950..149A, 2023ApJ...957L..28M}). 

\subsection{Disk Mass of AC Her}
\cite{2015A&A...578A..40H} found that the dust mass in the disk should be greater than $0.001\,{\rm M_\odot}$ in order to fit the millimeter wave fluxes emitted by the dust grains.
Assuming a conventional gas-to-dust ratio of $100$, this implies a total disk mass of at least $0.1\,{\rm M_\odot}$. 
However, we have shown that such a disk drives ZKL oscillations of the binary and therefore the disk will deviate from polar alignment (see the green lines in Figure~\ref{fig:ecc_inc_mass}). 
Even for a disk with a mass of $0.01\,{\rm M_\odot}$ (gas-to-dust ratio of $10$), the inclination of the disk relative to the binary still drops to $76^\circ$ within $2\times10^4$ years. 
Therefore, as \cite{2015A&A...578A..40H} mentioned, the gas-to-dust ratio of the disk should be much lower.

\begin{figure}
    \includegraphics[width=\columnwidth]{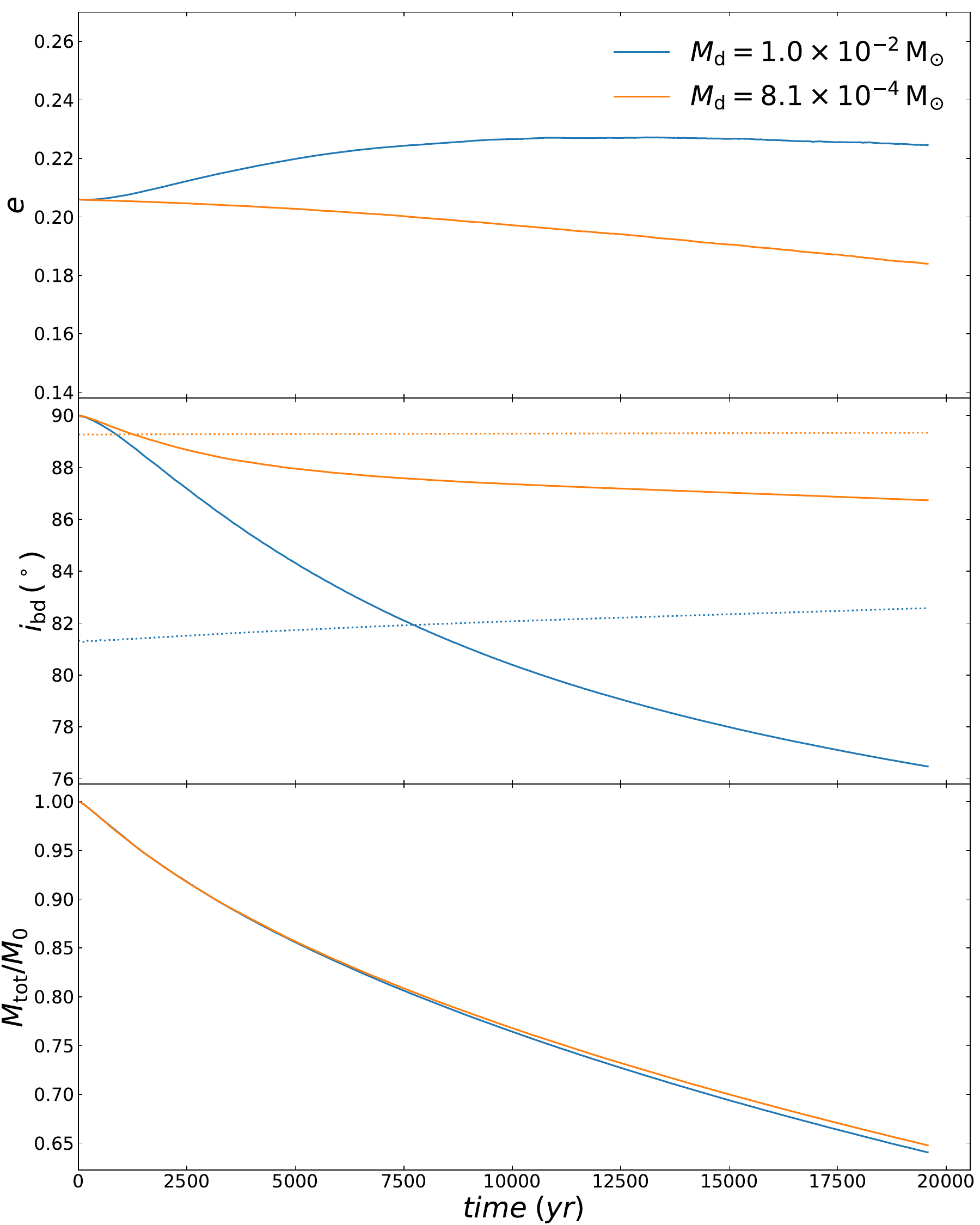}
    \caption{
             Eccentricities of the binary (upper panel), inclinations of the disk relative to the binary (bottom panel), and disk mass (bottom panel) as functions of time for simulation run~4 and 12.
             The blue and orange lines represent the initial mass of $M_{\rm d} = 0.01\,{\rm M_\odot}$ and $8.1\times10^{-4}\,{\rm M_\odot}$, respectively.
             The dotted lines in the middle panel show the disk's stationary inclination relative to the binary based on Equation~\ref{eq:StableInc}.
            }
    \label{fig:StableTest}
\end{figure}
Subsequently, \cite{2021A&A...648A..93G} determined the total disk mass to be $8.1\times10^{-4}\,{\rm M_\odot}$ by fitting a model to the CO emission from the disk.
To verify if this mass allows for a stable, polar-aligned disk, we perform a further simulation with an initial disk mass of $8.1\times10^{-4}\,{\rm M_\odot}$ (run 12 in Table~\ref{tab:simpara}).
In Figure~\ref{fig:StableTest}, we compare the evolution of the binary eccentricity and disk inclination with respect to the binary for this low mass case to those for the $M_{\rm d} = 0.01\,{\rm M_\odot}$ case.
We find that the inclination of the disk relative to the binary orbit drops by only a few degrees for the $M_{\rm d} = 8.1\times10^{-4}\,{\rm M_\odot}$ case. 
\cite{2023ApJ...957L..28M} indicated that the inclination of the disk relative to binary should be $96.5^{\circ}$ with an uncertainty of about $10^\circ$. 
Therefore, the disk's inclination with respect to the binary from our simulation falls within the observational error range.
Meanwhile, the binary's eccentricity does not change significantly until the end of our simulation.
Thus, the disk mass of $8.1\times10^{-4}\,{\rm M_\odot}$ is a plausible value for the AC Her system. 

\subsection{Disk extension of AC Her}
\begin{figure}
    \includegraphics[width=\columnwidth]{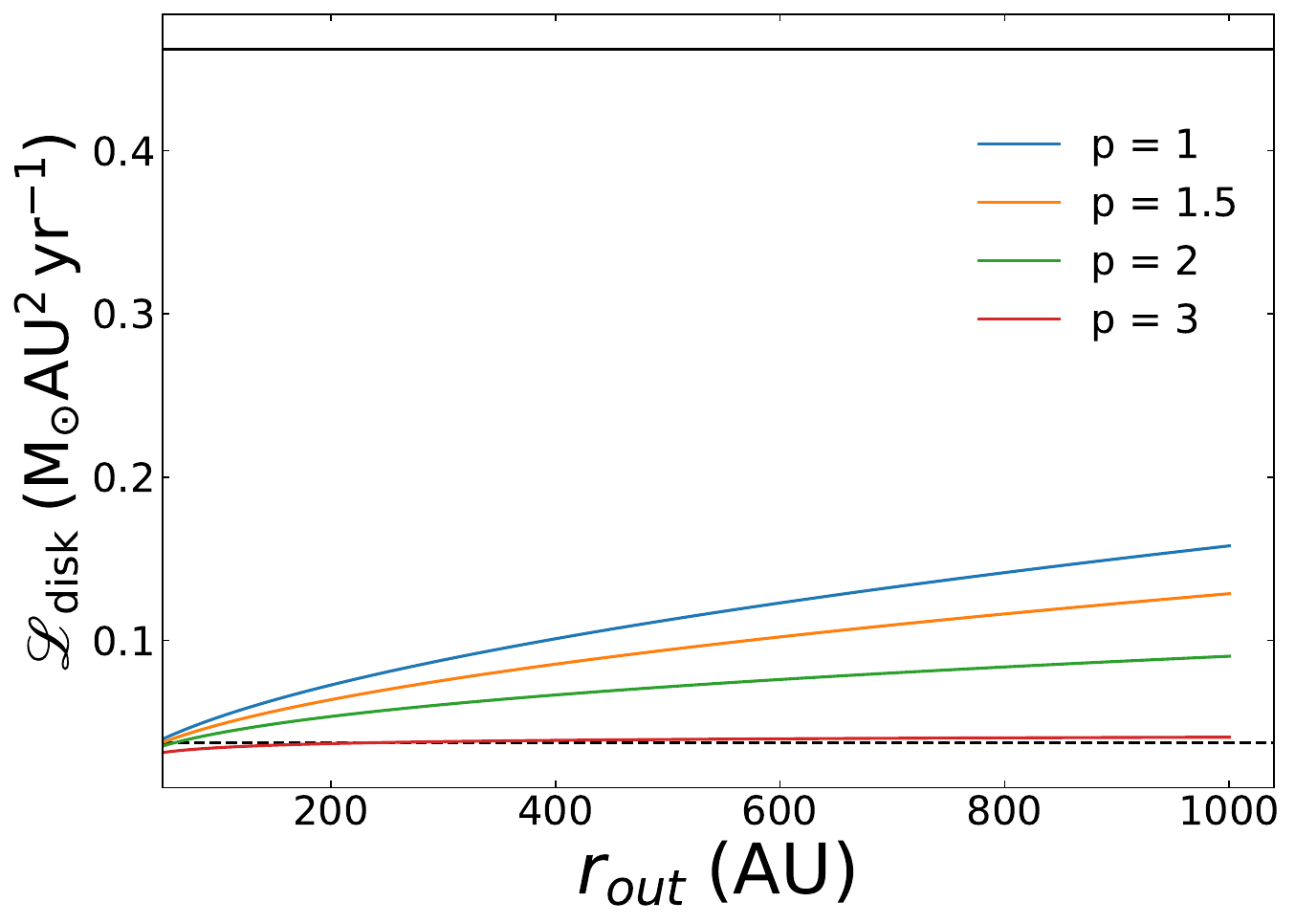}
    \caption{
             The initial setup angular momentum for the disk as a function of the extended radius. 
             The blue, orange, green, and red lines represent the power law index $p$ for the density profiles. 
             The black solid line and the black dashed line are the angular momentum for $M_{\rm d} = 0.01\,{\rm M_\odot}$ (run~4) and $M_{\rm d} = 8.1\times10^{-4}\,{\rm M_\odot}$ (run~12), respectively.
             }
    \label{fig:ldisk_vs_rout}
\end{figure}
Another point of contention is about the radial extent of the disk. 
\cite{2015A&A...578A..40H} reported the outer radius is about $200\,{\rm AU}$, while \cite{2021A&A...648A..93G} indicated that the disk could extend to $1000\,{\rm AU}$.
The total angular momentum could increase for a given disk mass if the disk extends further.
According to Equation~\ref{eq:StableInc}, which relates the stationary inclination to the disk's angular momentum, the total angular momentum significantly influences $i_{s}$ and, consequently, the disk's stability.
Therefore, it is worth examining how much the total disk's angular momentum will change when the disk mass is distributed differently. 
In Figure~\ref{fig:ldisk_vs_rout}, we show the total disk's angular momentum as a function of the outer radius with the same mass of $M_{\rm d} = 8.1\times10^{-4}\,{\rm M_\odot}$. 
The disk's surface density profile follows a power law with an index of $-p$ with an inner radius of $9\,{\rm AU}$. 

When $p$ is larger than $3$, we find that the total angular momentum $L_{\rm disk}$ is almost independent of $r_{\rm out}$. 
This is because a steep density profile with $p=3$ results in most of the disk mass being concentrated in the inner regions.
On the contrary, if $p$ is small, the density profile is relatively flat, and the outer part of the disk contributes more to the total angular momentum. 
As a result, when $r_{\rm out}$ extends from $50\,{\rm AU}$ to $1000\,{\rm AU}$, we found that the total angular momentum increases $2\sim 4$ times for $p$ values less than 2. 
However, even in these cases, the total angular momentum is still much smaller than that of the disk with $M_{\rm d} = 0.01\,{\rm M_\odot}$ (black solid line in Figure~\ref{fig:ldisk_vs_rout}). 
Therefore, the disk's extension has a smaller impact on the stability of polar alignment compared to its mass. 
In other words, it is possible for the disk to extend to $1000\,{\rm AU}$ as long as the total disk mass remains low.

\section{conclusions}\label{sec:conclusion}
Motivated by the recent observation of a polar-aligned disc in AC Her, we have examined the interaction between a post-AGB star binary and a polar-aligned circumbinary disk. 
Our findings indicate that large binary eccentricities can be driven by massive disks, potentially explaining some of the high eccentricities observed in post-AGB binaries.
Specifically, a disk with an initial mass of $0.1\,{\rm M_\odot}$ can efficiently enhance the binary eccentricity, reaching a maximum of 0.7.
Even when starting from a very low initial eccentricity (e.g., $e_0 = 0.01$), it can still grow, although on a longer timescale.
The binary eccentricity subsequently oscillates along with the disk inclination relative to the binary due to ZKL oscillations. 
These oscillations eventually cease due to viscous dissipation in the disk, resulting in a relatively large binary eccentricity that could be any value between the oscillation minimum and maximum. 

We also performed an evolution study for the disk mass. 
We found that the normalized mass loss rates are almost identical regardless of the initial disk mass, indicating that disks with larger masses lose mass more quickly. 
High-viscosity disks exhibit a rapid mass loss, leading to the cessation of oscillations before the eccentricity reaches its peak.
In contrast, mass loss occurs more slowly in low-viscosity disks, but these disks are prone to disk warping. 
However, it is challenging to determine the exact viscosity threshold below which a disk will break. 
Therefore, each case must be analyzed individually.

We explored the mass limit of the disk in the AC-Her system.
A large disk mass would be unstable for polar alignment and induce ZKL oscillation. 
Our simulation, based on an initial mass of $8.1\times 10^{-4}\,{\rm M_\odot}$ as suggested by \cite{2021A&A...648A..93G}, shows that this disk mass is preferable because the disk inclination relative to the binary remains stable within the observational error range proposed by \cite{2023ApJ...957L..28M}.
Furthermore, we found that the total angular momentum only increases by $2\sim 4$ times if the disk extends to $1000\,{\rm AU}$ from $50\,{\rm AU}$. 
Despite this change, the disk could still remain stable.
Thus, we conclude that the total disk mass should be on the order of $10^{-3}\,{\rm M_\odot}$ and that the disk could extend out to $1000\,{\rm AU}$ in the AC-Her system. 

\begin{acknowledgements}
We acknowledge support from NASA through grants 80NSSC21K0395 and 80NSSC19K0443.
\end{acknowledgements}


\bibliography{CBD}{}
\bibliographystyle{aasjournal}



\end{document}